\documentclass[10pt]{article}
\usepackage{latexsym,graphicx}
\newcommand{\be}{\begin{equation}}
\newcommand{\ee}{\end{equation}}
\def\n{\noindent}
\catcode `\@=11
\catcode `\@=12
\begin{document}
\begin{center}
\large{\bf {A New Class of String Cosmological Models in Cylindrically Symmetric 
Inhomogeneous Universe}} \\
\vspace{10mm}
\normalsize{Anirudh Pradhan \footnote{Corresponding author}, Mukesh Kumar Mishra $^2$ and Anil Kumar Yadav $^3$ }\\
\vspace{5mm}
\normalsize{$^{1,2}$ Department of Mathematics, Hindu Post-graduate College, 
Zamania-232 331, Ghazipur, India \\
$^1$E-mail : pradhan@iucaa.ernet.in \\
$^2$E-mail : mukeshkr.mishra@rediffmail.com} \\
\vspace{5mm}
\normalsize{$^{3}$Department of Physics, K. N. Govt. Post-graduate College, 
Sant Ravidas Nagar (Gyanpur), Bhadohi - 221 304, India \\
E-mail : abanilyadav@yahoo.co.in}\\ 
\vspace{5mm}
\end{center}
\vspace{10mm}
\begin{abstract} 
A new class of cylindrically symmetric inhomogeneous string cosmological models is 
investigated. To get the deterministic solution, it has been assumed that the 
expansion ($\theta$) in the model is proportional to the eigen value 
$\sigma^{1}~~_{1}$ of the shear tensor $\sigma^{i}~~_{j}$. The physical and geometric 
aspects of the model are also discussed.   
\end{abstract}
\smallskip
\n Keywords : String, Inhomogeneous universe, Cylindrical symmetry \\
\n PACS number: 98.80.Cq, 04.20.-q 
\section{Introduction}
In recent years, there has been considerable interest in string cosmology because 
cosmic strings play an important role in the study of the early universe. These strings 
arise during the phase transition after the big bang explosion as the temperature goes 
down below some critical temperature as predicted by grand unified theories (Zel'dovich 
et al., 1975; Kibble, 1976, 1980; Everett, 1981; Vilenkin, 1981). Moreover, the 
investigation of cosmic strings and their physical processes near such strings has 
received wide attention because it is believed that cosmic strings give rise to 
density perturbations which lead to formation of galaxies (Zel'dovich, 1980; 
Vilenkin, 1981). These cosmic strings have stress energy and couple to the 
gravitational field. Therefore, it is interesting to study the gravitational effect 
which arises from strings by using Einstein's equations. \\

The general treatment of strings was initiated by Letelier (1979, 1983) and 
Stachel (1980). Letelier (1979) obtained the general solution of Einstein's 
field equations for a cloud of strings with spherical, plane and a particular 
case of cylindrical symmetry. Letelier (1983) also obtained massive string 
cosmological models in Bianchi type-I and Kantowski-Sachs space-times. Benerjee 
et al. (1990) have investigated an axially symmetric Bianchi type I string dust 
cosmological model in presence and absence of magnetic field using a supplementary 
condition $\alpha = a \beta$ between metric potential where $\alpha = \alpha(t)$ and 
$\beta = \beta(t)$ and $a$ is constant. Exact solutions of string cosmology for 
Bianchi type-II, $-VI_{0}$, -VIII and -IX space-times have been studied by Krori et al. 
(1990) and Wang (2003). Wang (2004, 2005, 2006) has investigated bulk viscous string 
cosmological models in different space-times. Bali et al. (2001, 2003, 2005, 2006, 2007)
have obtained Bianchi type-I, -III, -V and type-IX string cosmological models in general 
relativity. The string cosmological models with a magnetic field are discussed by 
Chakraborty (1991), Tikekar and Patel (1992, 1994), Patel and Maharaj (1996). Ram 
and Singh (1995) obtained some new exact solution of string cosmology with and without a 
source free magnetic field for Bianchi type I space-time in the different basic form 
considered by Carminati and McIntosh (1980). Singh and Singh (1999) investigated string 
cosmological models with magnetic field in the context of space-time with $G_{3}$ 
symmetry. Singh (1995) has studied string cosmology with electromagnetic fields in 
Bianchi type-II, -VIII and -IX space-times. Lidsey, Wands and Copeland (2000) have 
reviewed aspects of super string cosmology with the emphasis on the cosmological 
implications of duality symmetries in the theory. Yavuz et al. (2005) have examined 
charged strange quark matter attached to the string cloud in the spherical symmetric 
space-time admitting one-parameter group of conformal motion. Recently Kaluza-Klein 
cosmological solutions are obtained by Yilmaz (2006) for quark matter attached to the 
string cloud in the context of general relativity. \\   

Cylindrically symmetric space-time play an important role in the study of the universe 
on a scale in which anisotropy and inhomogeneity are not ignored. Inhomogeneous 
cylindrically symmetric cosmological models have significant contribution in 
understanding some essential features of the universe such as the formation of 
galaxies during the early stages of their evolution. Bali and Tyagi (1989) and 
Pradhan et al. (2001, 2006) have investigated cylindrically symmetric inhomogeneous 
cosmological models in presence of electromagnetic field. Barrow and Kunze (1997, 1998) 
found a wide class of exact cylindrically symmetric flat and open inhomogeneous string 
universes. In their solutions all physical quantities depend on at most one space 
coordinate and the time. The case of cylindrical symmetry is natural because of the 
mathematical simplicity of the field equations whenever there exists a direction in 
which the pressure equal to energy density. \\

Recently Baysal et al. (2001), Kilinc and Yavuz (1996) have investigated 
some string cosmological models in cylindrically symmetric inhomogeneous universe. In 
this paper, we have revisited their solutions and obtained a new class of solutions. 
Here, we extend our understanding of inhomogeneous string cosmologies by investigating 
the simple models of non-linear cylindrically symmetric inhomogeneities outlined above. 
This paper is organized as follows: The metric and field equations are presented in 
Section $2$. In Section $3$, we deal with the solution of the field equations in three 
different cases. Finally, the results are discussed in Section $4$. The solutions 
obtained in this paper are new and different from the other author's solutions.  
\section{The Metric and Field  Equations}
We consider the metric in the form 
\begin{equation}
\label{eq1}
ds^{2} = A^{2}(dx^{2} - dt^{2}) + B^{2} dy^{2} + C^{2} dz^{2},
\end{equation}
where $A$, $B$ and $C$ are functions of $x$ and $t$.
The Einstein's field equations for a cloud of strings read as (Letelier, 1983) 
\begin{equation}
\label{eq2}
G^{j}_{i} \equiv  R^{j}_{i} - \frac{1}{2} R g^{j}_{i} = -(\rho u_{i}u^{j} - 
\lambda x_{i}x^{j}),
\end{equation}
where $u_{i}$ and $x_{i}$ satisfy conditions
\begin{equation}
\label{eq3}
u^{i} u_{i} = - x^{i} x_{i} = -1,
\end{equation}
and
\begin{equation}
\label{eq4}
u^{i} x_{i} = 0.
\end{equation}
Here, $\rho$ is the rest energy of the cloud of strings with massive particles 
attached to them. $\rho = \rho_{p} + \lambda$, $\rho_{p}$ being the rest energy 
density of particles attached to the strings and  $\lambda$ the density of tension 
that characterizes the strings. The unit space-like vector  $x^{i}$ represents 
the string direction in the cloud, i.e. the direction of anisotropy and the 
unit time-like vector  $u^{i}$ describes the four-velocity vector of the matter 
satisfying the following conditions 
\begin{equation}
\label{eq5}
g_{ij} u^{i} u^{j} = -1.
\end{equation}
In the present scenario, the comoving coordinates are taken as 
\begin{equation}
\label{eq6}
u^{i} = \left(0, 0, 0, \frac{1}{A}\right) 
\end{equation}
and choose $x^{i}$ parallel to x-axis so that
\begin{equation}
\label{eq7}
x^{i} = \left(\frac{1}{A}, 0, 0, 0 \right). 
\end{equation}
The Einstein's field equations (\ref{eq2}) for the line-element (\ref{eq1}) 
lead to the following system of equations:  
\[
G^{1}_{1} \equiv \frac{B_{44}}{B} + \frac{C_{44}}{C} - \frac{A_{4}}{A}
\left(\frac{B_{4}}{B} + \frac{C_{4}}{C}\right) - \frac{A_{1}}{A}
\left(\frac{B_{1}}{B} + \frac{C_{1}}{C}\right) -\frac{B_{1}C_{1}}{BC} + 
\frac{B_{4} C_{4}}{B C} 
\]
\begin{equation}
\label{eq8}
= \lambda A^{2},
\end{equation}
\begin{equation}
\label{eq9}
G^{2}_{2} \equiv \left(\frac{A_{4}}{A}\right)_{4} - \left(\frac{A_{1}}
{A}\right)_{1} + \frac{C_{44}}{C} - \frac{C_{11}}{ C} = 0,
\end{equation}
\begin{equation}
\label{eq10}
G^{3}_{3} \equiv \left(\frac{A_{4}}{A}\right)_{4} - \left(\frac{A_{1}}{A}\right)_{1} 
+ \frac{B_{44}}{B} - \frac{B_{11}}{B} =  0,
\end{equation}
\[
G^{4}_{4} \equiv - \frac{B_{11}}{B} - \frac{C_{11}}{C} + \frac{A_{1}}{A}
\left(\frac{B_{1}}{B} + \frac{C_{1}}{C}\right) + \frac{A_{4}}{A}\left(\frac{B_{4}}{B} 
+ \frac{C_{4}}{C}\right) - \frac{B_{1}C_{1}}{BC}  + \frac{B_{4} C_{4}}{B C} 
\]
\begin{equation}
\label{eq11}
= \rho A^{2},
\end{equation}
\begin{equation}
\label{eq12}
G^{1}_{4} \equiv \frac{B_{14}}{B} + \frac{C_{14}}{C} - \frac{A_{4}}{A}\left(\frac{B_{1}}{B}
 + \frac{C_{1}}{C}\right) - \frac{A_{1}}{A}\left(\frac{B_{4}}{B} + \frac{C_{4}}{C}\right) = 0,
\end{equation}
where the sub indices $1$ and $4$ in A, B, C and elsewhere denote differentiation
with respect to $x$ and $t$, respectively.

The velocity field $u^{i}$ is irrotational. The scalar expansion $\theta$, shear scalar 
$\sigma^{2}$, acceleration vector $\dot{u}_{i}$ and proper volume $V^{3}$ are respectively 
found to have the following expressions:
\begin{equation}
\label{eq13}
\theta = u^{i}_{;i} = \frac{1}{A}\left(\frac{A_{4}}{A} + \frac{B_{4}}{B} + \frac{C_{4}}{C}
\right),
\end{equation}
\begin{equation}
\label{eq14}
\sigma^{2} = \frac{1}{2} \sigma_{ij} \sigma^{ij} = \frac{1}{3}\theta^{2} - \frac{1}{A^{2}}
\left(\frac{A_{4}B_{4}}{AB} + \frac{B_{4}C_{4}}{BC} + \frac{C_{4}A_{4}}{CA}\right),
\end{equation}
\begin{equation}
\label{eq15}
\dot{u}_{i} = u_{i;j}u^{j} = \left(\frac{A_{1}}{A}, 0, 0, 0\right), 
\end{equation}
\begin{equation}
\label{eq16}
V^{3} = \sqrt{-g} = A^{2} B C,
\end{equation}
where $g$ is the determinant of the metric (\ref{eq1}). Using the field equations and 
the relations (\ref{eq13}) and (\ref{eq14}) one obtains the Raychaudhuri's equation as
\begin{equation}
\label{eq17}
\dot{\theta} = \dot{u}^{i}_{;i} - \frac{1}{3}\theta^{2} - 2 \sigma^{2} - \frac{1}{2} 
\rho_{p},
\end{equation}
where dot denotes differentiation with respect to $t$ and
\begin{equation}
\label{eq18}
R_{ij}u^{i}u^{j} = \frac{1}{2}\rho_{p}.
\end{equation}
 
With the help of equations (\ref{eq1}) - (\ref{eq7}), the Bianchi identity 
$\left(T^{ij}_{;j}\right)$ reduced to two equations:
\begin{equation}
\label{eq19}
\rho_{4} - \frac{A_{4}}{A}\lambda + \left(\frac{A_{4}}{A} + \frac{B_{4}}{B} + 
\frac{C_{4}}{C}\right)\rho = 0
\end{equation}
and
\begin{equation}
\label{eq20}
\lambda_{1} - \frac{A_{1}}{A}\rho + \left(\frac{A_{1}}{A} + \frac{B_{1}}{B} + 
\frac{C_{1}}{C}\right)\lambda = 0.
\end{equation}
Thus due to all the three (strong, weak and dominant) energy conditions, one finds 
$\rho \geq 0$ and $\rho_{p} \geq 0$, together with the fact that the sign of $\lambda$ 
is unrestricted, it may take values positive, negative or zero as well.  
\section{Solutions of the Field  Equations}

As in the case of general-relativistic cosmologies, the introduction of inhomogeneities 
into the string cosmological equations produces a considerable increase in mathematical 
difficulty: non-linear partial differential equations must now be solved. In practice, 
this means that we must proceed either by means of approximations which render the non-
linearities tractable, or we must introduce particular symmetries into the metric of the 
space-time in order to reduce the number of degrees of freedom which the inhomogeneities 
can exploit. \\

Here to get a determinate solution, let us assume that expansion ($\theta$) 
in the model is proportional to the eigen value $\sigma^{1}~~_{1}$ of  the shear tensor 
$\sigma^{i}~~_{j}$. This condition leads to
\begin{equation}
\label{eq21}
A = (BC)^{n},
\end{equation}
where $n$ is a constant. Equations (\ref{eq9}) and (\ref{eq10}) lead to
\begin{equation}
\label{eq22}
\frac{B_{44}}{B} - \frac{B_{11}}{B} = \frac{C_{44}}{C} - \frac{C_{11}}{C}.
\end{equation}
Using (\ref{eq21}) in (\ref{eq12}), yields 
\begin{equation}
\label{eq23}
\frac{B_{41}}{B} + \frac{C_{41}}{C} - 2n \left(\frac{B_{4}}{B} + 
\frac{C_{4}}{C}\right)\left(\frac{B_{1}}{B} + \frac{C_{1}}{C}\right) = 0.
\end{equation}
To find out deterministic solutions, we consider the following three cases: 
$$ (i) B = f(x)g(t) ~ ~ \mbox{and} ~ ~ C = h(x) k(t),$$
$$ (ii) B = f(x)g(t) ~ ~ \mbox{and} ~ ~ C = f(x) k(t),$$
$$ (iii) B = f(x)g(t) ~ ~ \mbox{and} ~ ~ C = h(x) g(t).$$
The two cases (i) and (ii) are discussed by Baysal et al. (2001) and the 
last case (iii) is discussed by Kilinc and Yavuz (1996). We revisit their 
solutions and obtain a new class of solutions for all these cases and discuss 
their consequences separately below in this paper. Our solutions are different 
from these author's solutions.
\subsection{Case(i):} $B = f(x)g(t)$ and $C = h(x)k(t)$ \\

In this case equation (\ref{eq23}) reduces to
\begin{equation}
\label{eq24}
\frac{f_{1}/f}{h_{1}/h} = - \frac{(2n - 1)(k_{4}/k) + 2n(g_{4}/g)}{(2n - 1)(g_{4}/g) + 
2n(k_{4}/k)} = K \mbox{(constant)},
\end{equation}
which leads to
\begin{equation}
\label{eq25}
\frac{f_{1}}{f} = K\frac{h_{1}}{h}
\end{equation}
and
\begin{equation}
\label{eq26}
\frac{k_{4}/k}{g_{4}/g} = \frac{K - 2nK - 2n}{2nK + 2n - 1} = a \mbox{(constant)}.
\end{equation}
From Eqs. (\ref{eq25}) and (\ref{eq26}), we obtain
\begin{equation}
\label{eq27}
f = \alpha h^{K}
\end{equation}
and
\begin{equation}
\label{eq28}
k = \delta g^{a},
\end{equation}
where $\alpha$ and $\delta$ are integrating constants. Eq. (\ref{eq22}) reduces to  
\begin{equation}
\label{eq29}
\frac{g_{44}}{g} - \frac{k_{44}}{k} = \frac{f_{11}}{f} - \frac{h_{11}}{h} = N,
\end{equation}
where $N$ is a constant. Using the functional values of B and C in (\ref{eq22}), 
we obtain 
\begin{equation}
\label{eq30}
g g_{44} + a g_{4}^{2} = - \frac{N}{a - 1}g^{2},
\end{equation}
which leads to 
\begin{equation}
\label{eq31}
g = \beta^{\frac{1}{a + 1}}\cosh^{\frac{1}{a + 1}}(bt + t_{0}),
\end{equation}
where $\beta$ and $t_{0}$ are constants of integration and 
$$b = \sqrt{a(a + 1)}. $$
Thus from Eq. (\ref{eq28}) we get
\begin{equation}
\label{eq32}
k = \delta \beta^{\frac{a}{a + 1}}\cosh^{\frac{a}{a + 1}}(bt + t_{0}).
\end{equation}
From Eqs. (\ref{eq25}) and (\ref{eq29}), we obtain 
\begin{equation}
\label{eq33}
h h_{11} + K h_{1}^{2} = \frac{N}{K - 1}h^{2},
\end{equation}
which leads to
\begin{equation}
\label{eq34}
h = \ell^{\frac{1}{K + 1}}\cosh^{\frac{1}{K + 1}}(rx + x_{0}),
\end{equation}
where $\ell$ and $x_{0}$ are constants of integration and 
$$ r = \sqrt{K(K + 1)}. $$
Hence from Eq. (\ref{eq27}) we have 
\begin{equation}
\label{eq35}
f = \alpha \ell^{\frac{K}{K + 1}}\cosh^{\frac{K}{K + 1}}(rx + x_{0}).
\end{equation}
It is worth mentioned here that equations (\ref{eq30}) and (\ref{eq33}) are 
fundamental basic differential equations for which we have reported new 
solutions given by equations (\ref{eq31}) and (\ref{eq34}). \\
  
Thus, we obtain 
\begin{equation}
\label{eq36}
B = fg = Q\cosh^{\frac{K}{K + 1}}(rx + x_{0})\cosh^{\frac{1}{a + 1}}(bt + t_{0}),
\end{equation}
\begin{equation}
\label{eq37}
C = hk = R\cosh^{\frac{1}{K + 1}}(rx + x_{0})\cosh^{\frac{a}{a + 1}}(bt + t_{0}),
\end{equation}
and
\begin{equation}
\label{eq38}
A = (BC)^{n} = M\cosh^{n}(rx + x_{0})\cosh^{n}(bt + t_{0}),
\end{equation}
where
$$ Q = \alpha \beta^{\frac{1}{a + 1}}\ell^{\frac{K}{K + 1}},$$
$$ R = \delta \beta^{\frac{a}{a + 1}}\ell^{\frac{1}{K + 1}},$$
$$ M = (QR)^{n}.$$
Hence the metric (\ref{eq1}) takes the form
\[
ds^{2}= M^{2}\cosh^{2n}(rx + x_{0})\cosh^{2n}(bt + t_{0}) (dx^{2} - dt^{2}) + 
\]
\[
Q^{2} \cosh^{\frac{2K}{K + 1}}(rx + x_{0})\cosh^{\frac{2}{a + 1}}(bt + t_{0})dy^{2} + 
\]
\begin{equation}
\label{eq39}
R^{2} \cosh^{\frac{2}{K + 1}}(rx + x_{0})\cosh^{\frac{2a}{a + 1}}(bt + t_{0})dz^{2}. 
\end{equation}
By using the following transformation
\[
rX = rx + x_{0},
\]
\[
Y = Q y,
\]
\[
Z = R z
\]
\begin{equation}
\label{eq40}
bT = bt + t_{0} 
\end{equation}
the metric (\ref{eq39}) reduces to
\[
ds^{2}= M^{2}\cosh^{2n}(r X) \cosh^{2n}(b T) (dX^{2} - dT^{2}) + 
\]
\begin{equation}
\label{eq41}
\cosh^{\frac{2K}{K + 1}}(r X)\cosh^{\frac{2}{a + 1}}(b T)dY^{2} +  
\cosh^{\frac{2}{K + 1}}(r X)\cosh^{\frac{2a}{a + 1}}(b T)dZ^{2}. 
\end{equation}

In this case the physical parameters, i.e. the energy density $(\rho)$, the 
string tension density $(\lambda)$, the particle density $(\rho_{p})$ and 
kinematical parameters, i.e. the scalar of expansion $(\theta)$, 
shear tensor $(\sigma)$, the acceleration vector $(\dot{u}_{i})$ and the proper 
volume $(V^{3})$ for the model (\ref{eq41}) are given by  
\begin{equation}
\label{eq42}
\rho = \frac{\left[b^{2}\left(n + \frac{a}{(a + 1)^{2}}\right) \tanh^{2n}(bT) + r^{2} 
\left(n + \frac{K}{(K + 1)^{2}}\right)\tanh^{2}(rX) - r^{2}\right]}{M^{2}\cosh^{2n}
(rX) \cosh^{2n}(bT)},
\end{equation}
\begin{equation}
\label{eq43}
\lambda = \frac{\left[- b^{2}\left(n + \frac{a}{(a + 1)^{2}}\right) \tanh^{2n}(bT) 
- r^{2}\left(n + \frac{K}{(K + 1)^{2}}\right)\tanh^{2}(rX) + b^{2}\right]}
{M^{2}\cosh^{2n}(rX) \cosh^{2n}(bT)},
\end{equation}
\begin{equation}
\label{eq44}
\rho_{p} = \frac{\left[2b^{2}\left(n + \frac{a}{(a + 1)^{2}}\right) \tanh^{2n}(bT) 
+ 2r^{2}\left(n + \frac{K}{(K + 1)^{2}}\right)\tanh^{2}(rX) -b^{2} - r^{2}\right]}
{M^{2}\cosh^{2n}(rX) \cosh^{2n}(bT)},
\end{equation}
\begin{equation}
\label{eq45}
\theta = \frac{b(n + 1)\tanh(bT)}{M\cosh^{n}(rX) \cosh^{n}(bT)},
\end{equation}
\begin{equation}
\label{eq46}
\sigma^{2} = \frac{b^{2}\tanh^{2}(bT)[(a + 1)^{2}(n^{2} - n + 1) - 3a]}
{3(a + 1)^{2}M^{2}\cosh^{2n}(rX) \cosh^{2n}(bT)},
\end{equation}
\begin{equation}
\label{eq47}
\dot{u}_{i} = \Bigl(nr \tanh(rX), 0, 0, 0\Bigr),
\end{equation}
\begin{equation}
\label{eq48}
V^{3} = \sqrt{-g} = \cosh^{2n + 1}(rX)\cosh^{2n + 1}(bT).
\end{equation}
From equations (\ref{eq45}) and (\ref{eq46}), we obtain
\begin{equation}
\label{eq49}
\frac{\sigma^{2}}{\theta^{2}} = \frac{(a + 1)^{2}(n^{2} - n + 1) - 3a}{3(n + 1)^{2}
(a + 1)^{2}} = \mbox{(constant)}.
\end{equation}
The models (\ref{eq41}) represents expanding, shearing and non-rotating 
universe. If we choose the suitable values of constants $K$ and $M$, we find that energy 
conditions $\rho \geq 0$, $\rho_{p} \geq 0$ are satisfied. Since $\frac{\sigma}{\theta}$ 
is constant throughout, hence the model does not approach isotropy. In this solution all 
physical and kinematical quantities depend on at most one space coordinate and the time.
\subsection{Case(ii):} $B = f(x)g(t)$ and $C = f(x)k(t)$ \\
In this case equation (\ref{eq23}) reduces to
\begin{equation}
\label{eq50}
(4n -1)\frac{f_{1}}{f}\left(\frac{g_{4}}{g} + \frac{k_{4}}{k}\right) = 0.
\end{equation}
The equation (\ref{eq50}) leads to three cases:
\[
(a) ~ ~ n = \frac{1}{4},
\]
\[
(b) ~ ~ \frac{f_{1}}{f} = 0,
\]
\[
(c) ~~ \frac{g_{4}}{g} + \frac{k_{4}}{k} = 0.
\]
The case (a) reduces the number of equation to four but, with five unknowns which 
requires additional assumption for a viable solution. In the case (b), the model 
turns to be a particular case to the Bianchi type-I model. Therefore we consider 
the case (c) only.\\

Using condition (c) in equation (\ref{eq22}) leads to 
\begin{equation}
\label{eq51}
\frac{g_{44}}{g} = \frac{k_{44}}{k}.
\end{equation}
By using condition (c) in (\ref{eq51}), we get
\begin{equation}
\label{eq52}
g = e^{LT}, ~ ~ ~ k = e^{-LT},
\end{equation}
where $T = t + \frac{t_{0}}{b}$, $t_{0}$, $b$,  and $L$ are constants. From equations 
(\ref{eq9}) or (\ref{eq10}) and (\ref{eq52}), we have
\begin{equation}
\label{eq53}
ff_{11} - \frac{2n}{2n + 1}f^{2}_{1} - \frac{L^{2}}{2n + 1}f^{2} = 0.
\end{equation}
Solving (\ref{eq53}), we obtain
\begin{equation}
\label{eq54}
f = \ell^{2n + 1}\cosh^{2n + 1}(M_{0}x + x_{0}),
\end{equation}
where 
$$ M_{0} = \frac{\sqrt{2n}}{2n + 1}.$$
and $\ell$ and $x_{0}$ are constants of integration. 

It is important to mention here that (\ref{eq53}) is the basic equation for which 
new solution is obtained as given by (\ref{eq54}). \\

Hence, we obtain
\begin{equation}
\label{eq55}
B = fg = Q_{0}e^{LT}\cosh^{2n + 1}(M_{0}x + x_{0})
\end{equation}
and
\begin{equation}
\label{eq56}
C = fk = Q_{0}e^{-LT}\cosh^{2n + 1}(M_{0}x + x_{0}), 
\end{equation}
where $Q_{0} = \ell^{2n + 1}$. Therefore
\begin{equation}
\label{eq57}
A = (BC)^{n} = N_{0} \cosh^{2n(2n + 1)}(M_{0}x + x_{0}),
\end{equation}
where $N_{0} = Q^{2n}_{0}$. \\

After suitable transformation of coordinates the metric (\ref{eq1}) 
reduces to the form 
\begin{equation}
\label{eq58}
ds^{2} = N^{2}_{0} \cosh^{2nm}(M_{0}x)(dX^{2} - dT^{2}) + Q^{2}_{0}\cosh^{m}(M_{0}X) 
(e^{2LT}dY^{2} + e^{-2LT}dZ^{2}),
\end{equation}
where $m = 2(2n + 1)$. \\

In this case the physical parameters $\rho$, $\lambda$, $\rho_{p}$ and kinematical 
parameters $\theta$, $\sigma$, $\dot{u}_{i}$ and $V^{3}$ for the model (\ref{eq58}) 
are given by
\begin{equation}
\label{eq59}
\rho = \frac{L^{2}}{N^{2}_{0}(2n + 1)\cosh^{2nm}(M_{0}X)}\left[(4n + 1)(2n - 1)
\tanh^{2}(M_{0}X) - (2n + 3)\right],
\end{equation}
\begin{equation}
\label{eq60}
\lambda = - \frac{L^{2}}{N_{0}^{2}\cosh^{2nm}(M_{0}X)}\left[(4n + 1)\tanh^{2}(M_{0}X) 
- 1 \right],
\end{equation}
\begin{equation}
\label{eq61}
\rho_{p} = \frac{4 L^{2}}{N^{2}_{0}(2n + 1)\cosh^{2nm}(M_{0}X)}\left[n(4n + 1)
\tanh^{2}(M_{0}X) - (n + 1)\right],
\end{equation}
\begin{equation}
\label{eq62}
\theta = 0,
\end{equation}
\begin{equation}
\label{eq63}
\sigma^{2} = \frac{L^{2}}{N_{0}^{2}\cosh^{2nm}(M_{0}X)},
\end{equation}
\begin{equation}
\label{eq64}
\dot{u}_{i} = \Bigl(n m M_{0} \tanh(M_{0}X), 0, 0, 0 \Bigr),
\end{equation}
\begin{equation}
\label{eq65}
V^{3} = \sqrt{-g} = Q_{0}^{m}\cosh^{m(2n + 1)}(M_{0}X).
\end{equation}
In this case the expansion $\theta$, in model (\ref{eq58}), is zero. With the help of 
physical and kinematical parameters, we can determine some physical and geometric features 
of the model. All kinematical quantities are independent of $T$. In general, the model 
represents non-expanding, non-rotating and shearing universe. The acceleration vector 
$\dot{u}$ is zero for $n = 0$, $n = - \frac{1}{2}$. Choosing suitable values for $n$, we find 
that energy conditions $\rho \geq 0$, $\rho_{p} \geq 0$ are satisfied. The solutions 
identically satisfy the Bianchi identities given by (\ref{eq19}) and (\ref{eq20}). In this 
solution all physical and kinematical quantities depend on at most one space coordinate.\\
\subsection{Case(iii): } $B = f(x)g(t)$ and $C = h(x)g(t)$ \\

In this case equation (\ref{eq23}) reduces to  
\begin{equation}
\label{eq66}
(4n - 1)\frac{g_{4}}{g}\left(\frac{f_{1}}{f} + \frac{h_{1}}{h}\right) = 0.
\end{equation}
The equation (\ref{eq66}) leads to three cases:
\[
(a) ~ ~ n = \frac{1}{4},
\]
\[
(b) ~ ~ \frac{g_{4}}{g} = 0,
\]
\[
(c) ~~ \frac{f_{1}}{f} + \frac{h_{1}}{h} = 0.
\]
The case (a) reduces the number of equation to four but, with five unknowns which 
requires additional assumption for a viable model. In the case (b), which infers a 
constant $g$ refers to the static solution. Therefore we consider the case 
\begin{equation}
\label{eq67}
\frac{f_{1}}{f} + \frac{h_{1}}{h} = 0.
\end{equation}
which produces non-static and physically meaningful solution as follows. 
Equation (\ref{eq67}) leads to
\begin{equation}
\label{eq68}
\frac{f_{11}}{f} = \frac{h_{11}}{h}.  
\end{equation}
Equation (\ref{eq68}), after integrating, gives
\begin{equation}
\label{eq69}
f = e^{L_{0} X}, ~ ~ ~ h = e^{-L_{0}X},
\end{equation}
where $X = x + \frac{x_{0}}{r}$ and $x_{0}$, $r$, $L_{0}$ are constants. From equations 
(\ref{eq9}) or (\ref{eq10}) and (\ref{eq69}), we have
\begin{equation}
\label{eq70}
g g_{44} - \frac{2n}{2n + 1}g^{2}_{4} - \frac{L^{2}_{0}}{2n + 1}g^{2} = 0,
\end{equation}
which after integration gives
\begin{equation}
\label{eq71}
g = \ell^{2n + 1}_{0}\cosh^{2n + 1}(K_{0}t + t_{0}),
\end{equation}
where $K_{0} = \frac{\sqrt{2n}}{2n + 1}$ and $\ell_{0}$, $t_{0}$ are constants of 
integration. It is important to mention here that (\ref{eq70}) is the basic equation 
for which new solution is obtained as given by \ref{eq71}). \\

Thus we obtain
\begin{equation}
\label{eq72}
B = fg = D e^{L_{0}X} \cosh^{2n + 1}(K_{0}t + t_{0})
\end{equation}
and
\begin{equation}
\label{eq73}
C = hg = D e^{-L_{0}X} \cosh^{2n + 1}(K_{0}t + t_{0}), 
\end{equation}
where $D = \ell^{2n + 1}$. Therefore 
\begin{equation}
\label{eq74}
A = (BC)^{n} = P \cosh^{2n(2n + 1)}(K_{0}t + t_{0}),
\end{equation}
where $P = D^{2n}$. \\

After suitable transformation of coordinates the metric (\ref{eq1}) reduces to 
the form
\begin{equation}
\label{eq75}
ds^{2} = P^{2}\cosh^{2ns}(K_{0}T)(dX^{2} - dT^{2}) + D^{2}\cosh^{s}(K_{0}T)
\left(e^{2L_{0}X}dY^{2} + e^{- 2L_{0}X}dZ^{2}\right),
\end{equation}
where $s = 2(2n + 1)$. \\

In this case the physical parameters $\rho$, $\lambda$, $\rho_{p}$ and kinematical 
parameters $\theta$, $\sigma$, $\dot{u}_{i}$ and $V^{3}$ for the model (\ref{eq75}) 
are given by
\begin{equation}
\label{eq76}
\rho = \frac{K^{2}_{0}(4n + 1)(2n + 1)^{2}\tanh^{2}(K_{0}T)    - L^{2}_{0}}
{P^{2} \cosh^{2ns}(K_{0}T)},
\end{equation}
\begin{equation}
\label{eq77}
\lambda = \frac{K^{2}_{0}[(4n + 2) - (4n + 1)(4n^{2} - 1)\tanh^{2}(K_{0}T)] 
+ L^{2}_{0}}{P^{2} \cosh^{2ns}(K_{0}T)},
\end{equation}
\begin{equation}
\label{eq78}
\rho_{p} = \frac{K^{2}_{0}[4n(2n + 1)(4n + 1)\tanh^{2}(K_{0}T) - (4n + 2)] 
- 2L^{2}_{0}}{P^{2} \cosh^{2ns}(K_{0}T)},
\end{equation}
\begin{equation}
\label{eq79}
\theta = \frac{K_{0}s(n + 1)\tanh(K_{0}T)}{P \cosh^{ns}(K_{0}T)},
\end{equation}
\begin{equation}
\label{eq80}
\sigma^{2} = \frac{K^{2}_{0}s^{2}(2n - 1)^{2}\tanh^{2}(K_{0}T)}{12 P^{2} 
\cosh^{2ns}(K_{0}T)},
\end{equation}
\begin{equation}
\label{eq81}
\dot{u}_{i} = (0, 0, 0, 0)
\end{equation}
\begin{equation}
\label{eq82}
V^{3} = \sqrt{-g} = (PD)^{2} \cosh^{(2n + 1)s}(K_{0}T).
\end{equation}
Therefore
\begin{equation}
\label{eq83}
\frac{\sigma^{2}}{\theta^{2}} = \frac{4n^{2} - 4n + 1}{12(n + 1)^{2}} = \mbox{constant}.
\end{equation}
The models (\ref{eq75}) represents an expanding, shearing and non-rotating 
universe. We find that energy conditions 
$\rho \geq 0$, $\rho_{p} \geq 0$ are satisfied if we choose $n > -\frac{1}{4}$ and $L_{0} 
\ne 0$ and we get physically significant string cosmology model. The energy density 
$\rho$, the string tension density $\lambda$ and particle density $\rho_{p}$ at all finite 
spatial location tend to constant value as $T \to 0$. Since we observe that $\frac{\sigma}
{\theta}$ is constant throughout, hence the model does not approach isotropy. The 
solutions identically satisfy the Bianchi identities given by (\ref{eq19}) and 
(\ref{eq20}). In this solution all physical and kinematical quantities depend on at 
most one time coordinate.
\section{Concluding Remarks}
In the study, we have presented a new class of exact solutions of Einstein's field 
equations for inhomogeneous cylindrically symmetric space-time with string sources 
which are different from the other author's solutions. In these solutions all physical 
quantities depend on at most one space coordinate and the time.

In case (i), the models (\ref{eq41}) represents an expanding, shearing and non-rotating 
universe and all physical and kinematical parameters depend on at most one space coordinate 
and the time.

In case (ii), the model (\ref{eq58}) represents non-expanding, non-rotating and shearing universe. 
The solutions identically satisfy the Bianchi identities given by (\ref{eq19}) and (\ref{eq20}). In this 
solution all physical and kinematical quantities depend on at most one space coordinate.

 In case (iii), the models (\ref{eq75}) represents expanding, shearing and non-rotating 
universe. The solutions identically satisfy the Bianchi identities given by (\ref{eq19}) and 
(\ref{eq20}). In this solution all physical and kinematical quantities depend on at 
most one time coordinate.

\section*{Acknowledgements} 
One of the Authors (A. P.) would like to thank Professor G. Date, IMSc., Chennai, India 
for providing facility where part of this work was carried out. 

\end{document}